\documentclass[aps,pre,reprint,groupedaddress,amsmath,amssymb,superscriptaddress,floatfix]{revtex4-2}

\usepackage{hyperref}
\usepackage{graphicx}
\usepackage[capitalize]{cleveref}
\usepackage[svgnames]{xcolor}

\usepackage{siunitx}[=v2]
\DeclareSIUnit\bit{\textrm{bit}}

\usepackage[version=4]{mhchem}
\usepackage{bm}

\DeclareMathOperator{\prob}{\mathrm{P}}

\renewcommand{\vec}{\bm}
\newcommand{\mat}{\bm}

\newcommand{\avg}[1]{\left\langle#1\right\rangle}
\newcommand{\f}[2][]{\tilde#2(#1\omega)} 
\newcommand{\spec}[2][]{S_{#2}(#1\omega)} 

\newcommand{\dg}{\tilde g}
\newcommand{\freqnoise}[1][]{|N_{#1}(\omega)|^2}
\newcommand{\freqgain}[1][]{|K_{#1}(\omega)|^2}

\begin{document}

\title{Mutual Information Rate --- Linear Noise Approximation and Exact Computation}

\author{Manuel Reinhardt}
\thanks{These authors contributed equally}
\affiliation{AMOLF, Science Park 104, 1098 XG Amsterdam, The Netherlands}
\author{Age J. Tjalma}
\thanks{These authors contributed equally}
\affiliation{AMOLF, Science Park 104, 1098 XG Amsterdam, The Netherlands}
\author{Anne-Lena Moor}
\affiliation{Max Planck Institute of Molecular Cell Biology and Genetics, 01307 Dresden, Germany}
\affiliation{Center for Systems Biology Dresden, 01307 Dresden, Germany}
\author{Christoph Zechner}
\affiliation{Scuola Internazionale Superiore di Studi Avanzati, 34136 Trieste, Italy}
\affiliation{Max Planck Institute of Molecular Cell Biology and Genetics, 01307 Dresden, Germany}
\affiliation{Center for Systems Biology Dresden, 01307 Dresden, Germany}
\author{Pieter Rein ten Wolde}%
\email{tenwolde@amolf.nl}
\affiliation{AMOLF, Science Park 104, 1098 XG Amsterdam, The Netherlands}

\date{August 29, 2025}

\begin{abstract}
Efficient information processing is crucial for both living organisms and engineered systems. The mutual information rate, a core concept of information theory, quantifies the amount of information shared between the trajectories of input and output signals, and enables the quantification of information flow in dynamic systems. 
A common approach for estimating the mutual information rate is the Gaussian approximation which assumes that the input and output trajectories follow Gaussian statistics. However, this method is limited to linear systems, and its accuracy in nonlinear or discrete systems remains unclear.
In this work, we assess the accuracy of the Gaussian approximation for non-Gaussian systems by leveraging Path Weight Sampling (PWS), a recent technique for exactly computing the mutual information rate. 
In two case studies, we examine the limitations of the Gaussian approximation.
First, we focus on discrete linear systems and demonstrate that, even when the system’s statistics are nearly Gaussian, the Gaussian approximation fails to accurately estimate the mutual information rate.
Second, we explore a continuous diffusive system with a nonlinear transfer function, revealing significant deviations between the Gaussian approximation and the exact mutual information rate as nonlinearity increases.
Our results provide a quantitative evaluation of the Gaussian approximation’s performance across different stochastic models and highlight when more computationally intensive methods, such as PWS, are necessary. 
\end{abstract}

\maketitle

\section{\label{sec:intro}Introduction}
For the functioning of both living and engineered systems 
it is paramount that they collect and process information effectively. Increasingly, it has become clear that beyond instantaneous properties, the dynamic features of an input signal or system output often encode valuable information \cite{2009.Tostevin, 2010.Tostevin, 2021.Mattinglypd, 2023.Moor, 2023.Reinhardt, 2023.Hahn}. 
Prime examples in biology include bacterial chemotaxis, which responds to temporal changes in concentration \cite{segall1986temporal},
the transcription factor NF-$\kappa$B, which encodes information about input signals in its dynamic response \cite{covert2005achieving}, 
and neuronal information processing, where information is encoded in the sequence and timing of spikes \cite{strong1998application}. 
Beyond biology, dynamic input signals are critical for various sensing systems, such as those used in automated factories or self-driving cars.

To understand and evaluate the performance, potential improvements, and limitations of these systems in processing information, we need appropriate metrics that capture their full information processing capability. 
Information theory, introduced by \citet{1948.Shannon}, provides the most general mathematical framework for such metrics.
The mutual information measures how much one random variable reduces uncertainty about another, quantified in bits. 
It is relatively straightforward to quantify the information shared between scalar properties of the input and output, as has been done in various forms \cite{ziv2007optimal, cheong2011information, dubuis2013positional, palmer2015predictive, chalk2018toward, bauer2021trading, sachdeva2021optimal, tjalma2023trade, tjalma2024predicting}. 
However, capturing all information in dynamical properties of the input and the output is much more challenging.
To do so, one must consider the information encoded in time-varying trajectories of the variables of interest. 
Yet, due to the high dimensionality of the trajectory space, computing the mutual information between such trajectories is notoriously difficult.

A major advance for biological systems has been the Gaussian approximation of the mutual information rate \cite{2009.Tostevin, 2010.Tostevin}, based on the assumption of input and output trajectories following jointly Gaussian statistics.  
This assumption makes it possible to compute the mutual information rate  directly from the two-point correlation functions of the input and output.
It is thus straightforward to apply the Gaussian approximation to experimental data.
Moreover, given a mechanistic model of the underlying dynamics, the Gaussian approximation can be used to derive analytical expressions for the information rate \cite{2009.Tostevin, 2010.Tostevin, 2021.Mattinglypd}.
Crucially, however, the assumption of Gaussian statistics restricts the method to linear systems with additive Gaussian noise \cite{2006.Cover}.

Understanding when the Gaussian approximation is accurate is critical because many real-world systems, such as biological and engineered sensory systems, exhibit nonlinear dynamics. 
This includes features such as bimodality, discrete jumps, or heavy tails, which deviate from Gaussian dynamics.
Such non-Gaussian behavior typically results from intrinsic nonlinearities in the system.
Determining the degree of a system’s deviation from linearity is difficult \cite{2008.Tkacik,deronde2010effect,mitra2001nonlinear}, and
the extent to which the approximation loses accuracy in nonlinear systems is unclear.
Thus, although the Gaussian approximation offers a simple method to estimate information transmission, it remains an open question under what conditions this approximation is sufficiently accurate.

Until recently, addressing this question has been hard because there was no reliable benchmark for the exact information rate.
Without a method to compute the true information rate of a non-Gaussian system, it is impossible to rigorously assess the accuracy of the Gaussian approximation.
This gap was filled recently by the development of two independent methods \cite{2023.Reinhardt, 2023.Moor} for computing the information rate accurately even in systems that significantly deviate from Gaussian behavior. 
Here we leverage one of these methods: Path Weight Sampling (PWS) \cite{2023.Reinhardt}. This Monte Carlo technique is an exact method for calculating the mutual information rate for a wide range of stochastic models.
Using PWS calculations, we can directly evaluate the accuracy of the Gaussian approximation in models that exhibit explicit non-Gaussian features, and study the approximation's robustness in typical applications.

In this article, we probe the accuracy of the Gaussian approximation for the information rate in two case studies.
The first focuses on Markov jump processes, where the statistics are non-Gaussian due to their discrete nature. We first show that the Gaussian approximation fails to accurately estimate the mutual information rate in this case, even in the regime where the statistics are nearly Gaussian \cite{2023.Moor,2023.Reinhardt}. We then apply a recently developed ``discrete approximation'' \cite{2023.Moor} and show it to be much more accurate.
We briefly discuss this surprising failure of the Gaussian approximation, which has been more extensively analyzed in Ref.~\cite{2025.Moor}. Finally, we still find regimes where neither of the approximations are highly accurate.

The second case study examines a continuous diffusive process with a nonlinear transfer function. We demonstrate how intrinsic nonlinearity can cause significant deviations between the Gaussian approximation and the true mutual information rate. By varying the degree of nonlinearity as well as the system's response time, we provide a comprehensive quantitative understanding of the Gaussian approximation’s limitations in nonlinear systems.
Additionally, we show that for such systems, the Gaussian approximation differs significantly when derived from empirical correlation functions compared to when it is analytically obtained from the nonlinear model, highlighting that the correct application of the approximation is important.

Our work translates into concrete recommendations on when to use which method for the computation of the information rate. It therefore enables researchers to more confidently determine when a simpler approximate method is sufficient, or when a more sophisticated method like PWS \cite{2023.Reinhardt} or the method developed by \citet{2023.Moor} should be used.

\section{Methods}

\subsection{The mutual information rate}

The mutual information between two random variables $S$ and $X$ is defined as
\begin{equation}
I(S, X) = \iint \prob(s, x) \ln \frac{\prob(s, x)}{\prob(s)\prob(x)}\ ds\,dx \,,
\end{equation}
or, equivalently, using Shannon entropies
\begin{equation}
\begin{aligned}
I(S, X) 
&= H(S) + H(X) - H(S, X) \\
&= H(S) - H(S|X) \\
&= H(X) - H(X|S)\,.
\end{aligned}
\end{equation}
In the context of a noisy communication channel, $S$ and $X$ represent the messages at the sending and receiving end, respectively. Then, $I(S, X)$ is the amount of information about $S$ that is communicated when only $X$ is received. If $S$ can be perfectly reconstructed from $X$, the mutual information reaches its maximal value $I(S,X)=H(S)$. On the contrary, if $S$ and $X$ are independent, $I(S, X)=0$. The mutual information thus is always non-negative and quantifies the degree of statistical dependence between two random variables.

For systems that continuously transmit information over time, this concept must be extended to trajectories $\vec{S}_T=\{S(t)\mid t \in [0, T]\}$ and $\vec{X}_T=\{X(t)\mid t \in [0, T]\}$. 
The mutual information between trajectories is defined analogously as
\begin{equation}
I(\vec{S}_T,\vec{X}_T) = \left\langle \ln\frac{\prob(\vec{s}_T, \vec{x}_T)}{\prob(\vec{s}_T) \prob(\vec{x}_T)} \right\rangle
\label{eq:trajectory_mi}
\end{equation}
where the expected value is taken with respect to the full joint probability of both trajectories. This quantity can be interpreted as the total information that is communicated over the time interval $[0, T]$.

Note that the total amount of information communicated over the time-interval $[0, T]$ is not directly related to the instantaneous mutual information $I(S(t), X(t))$ at any instant $0 \leq t \leq T$.
This is because auto-correlations within the input or output sequences reduce the amount of new information transmitted in subsequent measurements. 
Moreover, information can be encoded in temporal features of the trajectories, which cannot be captured by an instantaneous information measure.
Therefore, as previously pointed out \cite{2021.Meijers,2024.Fan}, the instantaneous mutual information $I(S(t), X(t))$ for any given $t$ does not provide a meaningful measure of information transmission.
To correctly quantify the amount of information transmitted per unit time, we must consider entire trajectories.

For that reason, the \emph{mutual information rate} is defined via the trajectory mutual information.
Let the input and output of a system be given by two continuous-time stochastic processes $\mathcal{S}=\{S(t)\mid t\in\mathbb{R}\}$ and $\mathcal{X}=\{X(t)\mid t\in\mathbb{R}\}$.
Then, for stationary systems, the mutual information rate between $\mathcal{S}$ and $\mathcal{X}$ is
\begin{equation}
\label{eq:ratedef}
R(\mathcal{S}, \mathcal{X}) = \lim_{T\to\infty} \frac{1}{T} I(\vec{S}_T, \vec{X}_T)\,,
\end{equation}
and quantifies the amount of information that can reliably be transmitted per unit time. The mutual information rate therefore represents an excellent performance measure for information processing systems.

In summary, the mutual information rate is \emph{the} crucial performance metric for stochastic information processing systems. However, its information-theoretic definition does not translate into an obvious scheme for computing it. As a result, various methods have been developed to compute or approximate the mutual information rate.

\subsection{Gaussian Approximation}

To significantly simplify the computation of the information rate it is often assumed that the input and output trajectories obey stationary Gaussian statistics. 
Under this assumption \cref{eq:trajectory_mi} simplifies to
\begin{equation}
I(\vec{S}_T,\vec{X}_T) = \frac{1}{2}\ln\frac{|\mat{C}_{ss}||\mat{C}_{xx}|}{|\mat{Z}|}, 
\end{equation}
where $|\mat{C}_{ss}|$ and $|\mat{C}_{xx}|$ are the determinants of the covariance matrices of the respective trajectories $\vec{S}_T$ and $\vec{X}_T$, and 
\begin{equation}
    \mat Z = \begin{pmatrix}
        \mat{C}_{ss} & \mat{C}_{sx} \\
        \mat{C}_{xs} & \mat{C}_{xx}
    \end{pmatrix}
\end{equation}
is the covariance matrix of their joint distribution. 

In the limit that the trajectory length $N=T/\Delta$, with the discretization $\Delta$, becomes infinitely long ($N\to\infty$) and continuous ($\Delta \to 0$), the information rate as defined in \cref{eq:ratedef} can be expressed in terms of the power spectral densities, or power spectra, of the Gaussian processes $\mathcal S$ and $\mathcal X$ \cite{2009.Tostevin, 2010.Tostevin}:
\begin{equation}
\label{eq:gaussdef}
    R(\mathcal{S}, \mathcal{X}) = -\frac{1}{4\pi}\int_{-\infty}^{\infty} d\omega \ln\left(1-\frac{|\spec{sx}|^2}{\spec{ss}\spec{xx}}\right).
\end{equation}
Here, $\spec{ss}$ and $\spec{xx}$ respectively are the power spectra of trajectories generated by $\mathcal S$ and $\mathcal X$, and $\spec{sx}$ is their cross-spectrum. The fraction 
\begin{equation}
    \phi_{sx}(\omega) = \frac{|\spec{sx}|^2}{\spec{ss}\spec{xx}}
\end{equation}
is known as the coherence, describing the distribution of power transfer between $\mathcal S$ and $\mathcal X$ over the frequency $\omega$. 

For systems that are neither Gaussian nor linear, there are two ways to still obtain an approximate Gaussian information rate. The first is to directly measure two-point correlation functions from data or simulations, and use these to numerically obtain the power spectra in \cref{eq:gaussdef}. The second is to use \citeauthor{vanKampen1992}'s linear noise approximation (LNA) and approximate the dynamics of the system to first order around a fixed point \cite{vanKampen1992} which makes it possible to analytically obtain approximate power spectra, see also \cref{app:gauss}. In this work, we will analyze both of these methods.

\subsection{Path Weight Sampling for diffusive systems}

To evaluate the accuracy of the Gaussian information rate for non-Gaussian systems, an exact method for determining the true information rate is required. Recently, a method called Path Weight Sampling (PWS) was developed, which computes the exact mutual information rate using Monte Carlo techniques without relying on approximations \cite{2023.Reinhardt}.

In Ref.~\cite{2023.Reinhardt}, PWS was introduced as a computational framework for calculating the mutual information rate in systems governed by master equations. 
Master equations provide an exact stochastic description of continuous-time processes with discrete state spaces, commonly used in models ranging from biochemical signaling networks to population dynamics.
However, many systems are not described by discrete state spaces and instead require a stochastic description based on diffusion processes or other stochastic models.
Fortunately, PWS is not restricted to systems described by master equations and can be extended to a variety of stochastic models. 

In general, PWS can be applied to any system that meets the following conditions: 
(i)~sampling from the input distribution $\prob(\vec{s}_T)$ is straightforward, 
(ii)~sampling from the conditional output distribution $\prob(\vec{x}_T \mid \vec{s}_T)$ is straightforward, and 
(iii)~the logarithm of the conditional probability density $\ln\prob(\vec{x}_T \mid \vec{s}_T)$, referred to as the path weight, can be evaluated efficiently.
For any stochastic model that satisfies these three criteria, the PWS  computation proceeds similarly to systems governed by master equations.

Briefly, PWS computes the trajectory Mutual Information using a Monte Carlo estimate of \cref{eq:trajectory_mi}
\begin{equation}
    \frac{\sum^N_{i=1}\left[ \ln \prob\left(\vec{x}^i_T \Bigm| \vec{s}^i_T\right) - \ln\prob\left(\vec{x}^i_T\right) \right] }{N}
    \label{eq:pws_monte_carlo}
\end{equation}
where $\vec{s}^1_T, \ldots, \vec{s}^N_T$ are independently drawn from $\prob(\vec{s}_T)$, and each $\vec{x}^i_T$ is drawn from $\prob(\vec{x}_T \mid \vec{s}^i_T)$.
As $N\to\infty$, this expression converges to the mutual information $I(\vec{S}_T,\vec{X}_T)$.
In \cref{eq:pws_monte_carlo}, the term $\ln\prob(\vec{x}_T \mid \vec{s}_T)$ can be evaluated directly (per criterion iii), but the marginal probability $\prob(\vec{x}_T)$ has to be computed separately for each output trajectory $\vec{x}^i_T$.
Typically, this has to be done numerically via marginalization, i.e., by computing the path integral
\begin{equation}
    \prob(\vec{x}_T) = \int d\vec{s}_T\ \prob(\vec{s}_T) \prob(\vec{x}_T \mid \vec{s}_T)
\end{equation}
using Monte Carlo integration.
Evaluating the marginalization integral efficiently is essential for computing the mutual information using PWS and discussed in detail in Ref.~\cite{2023.Reinhardt}.
In summary, PWS is a generic framework that can be used beyond systems defined by a master equation as long as a suitable generative model satisfying the three conditions above is available.

For this study, we extended PWS to compute the mutual information rate for systems with diffusive dynamics, described by Langevin equations. 
For such systems, the aforementioned conditions are inherently fulfilled and PWS can be applied.
Specifically, in a Langevin system, both the input $S(t)$ and the output $X(t)$ are stochastic processes given by the solution to a stochastic differential equation (SDE).
Using stochastic integration schemes like the Euler-Mayurama method, we can straightforwardly generate realizations $s(t)$ and $x(t)$ from the corresponding stochastic process.
These realizations are naturally time-discretized with the integration time step $\Delta t$.
For a time-discretized trajectory $\vec{x}=(x_1,\ldots,x_n)$, the path weight $\ln\prob(\vec{x} | \vec{s})$ is---up to a Gaussian normalization constant---given by the Onsager-Machlup action \cite{1953.Onsager}
\begin{equation}
    \ln\prob(\vec{x} \mid \vec{s}) =
    -\sum^{n-1}_{i=1} 
    \frac{1}{2\Delta t} 
    \left( \frac{\Delta x_i - v_i \Delta t}{\sigma(x_i)} \right)^2 + \text{const}
\end{equation}
where we used $\Delta x_i = x_{i+1} - x_i$, and $v_i = f(x_i, s_i)$ is the deterministic drift, and $\sigma(x_i)$ represents the white noise amplitude.
This expression captures the likelihood of a particular trajectory, given the stochastic dynamics of the system, and serves as the path weight in the PWS computation.

\section{Case Studies}

To investigate the conditions under which the Gaussian approximation deviates from the exact mutual information rate, we conducted two case studies.
In both studies we compare the Gaussian approximation against the exact mutual information rate, computed via PWS.
In the first case study we focus on a discrete linear system which is inspired by minimal motifs of cellular signaling.

\subsection{Discrete reaction system}
\label{sec:linsys}

We consider a simple linear reaction system of two species, $S$ and $X$, whose dynamics are governed by 4~reactions
\begin{align}
    \label{eq:reac1}\ce{\emptyset &->[\kappa] S} \\ 
    \label{eq:reac2}\ce{S &->[\lambda] \emptyset} \\ 
    \label{eq:reac3}\ce{S &->[\rho] S + X} \\ 
    \label{eq:reac4}\ce{X &->[\mu] \emptyset}\,. 
\end{align}
The reaction system is linear because each reaction has at most one reactant.
The trajectories of $S$ and $X$ are correlated because the production rate of $X$ depends on the copy number of $S$, and therefore information is transferred from $S$ to $X$.
This set of reactions can be interpreted as a simple motif for gene expression where $S$ is a transcription factor and $X$ represents the expressed protein.
In steady state, the mean copy numbers are given by $\bar{s} = \kappa\lambda^{-1}$ and $\bar{x} = \bar{s}\rho\mu^{-1}$.

The exact stochastic dynamics of this reaction system can be expressed by the chemical master equation \cite{vanKampen1992}. 
This equation describes the time-evolution of the probability distribution over the possible copy numbers of species $S$ and $X$, capturing the noise from the discrete chemical reaction events.
From this description, we can obtain the mutual information rate from $S$ to $X$ without approximations using PWS \cite{2023.Reinhardt}.

While the chemical master equation is an exact representation of the reaction system, for large copy numbers the stochastic dynamics are well approximated by a Gaussian model. The resulting Langevin equations can be systematically derived from the master equation using the LNA which yields
\begin{align}
    \label{eq:input-dynamics}\dot{s}(t) &= \kappa - \lambda  s(t) + \eta_s(t) \\ 
    \dot {x}(t) &= \rho s(t) - \mu x(t) + \eta_x(t)
    \label{eq:output-dynamics}
\end{align}
where $s$ and $x$ are continuous variables representing the copy numbers of $S$ and $X$, and $\eta_s, \eta_x$ are independent delta-correlated white noise terms with $\langle \eta^2_s \rangle = 2\lambda\bar{s}$ and $\langle \eta^2_x \rangle = 2\mu\bar{x}$, see \cref{app:gauss}.

The Gaussian approximation of the mutual information rate is derived from the LNA description.
Using this framework, 
\citet{2009.Tostevin} found an analytical expression for the mutual information rate of the motif in units of nats $s^{-1}$:
\begin{equation}
    R_\text{Gaussian} = \frac{\lambda}{2} \left( \sqrt{1 + \frac{\rho}{\lambda}} - 1 \right) \,.
    \label{eq:tostevin}
\end{equation}

More recently, \citet{2023.Moor} have derived a different expression for the mutual information rate of this reaction system by analytically approximating the relevant filtering equation, which is derived from the master equation, thus recognizing the discreteness of the state space. 
This approach explicitly differentiates the contributions of individual reactions to the noise amplitude of each component, while the LNA lumps their contributions together.
As we will discuss in more detail below, separately accounting for the noise from each reaction better captures the information transmitted via discrete systems, making this ``discrete approximation'' more accurate than the Gaussian approximation for this case study.
Nevertheless, the result is still based on an approximation that is only expected to be accurate for large copy numbers.
The expression for the mutual information rate in the discrete approximation is remarkably similar to the expression obtained using the Gaussian framework:
\begin{equation}
    R_\text{discrete} = \frac{\lambda}{2} \left( \sqrt{1 + 2\ \frac{\rho}{\lambda}} - 1 \right)
    \label{eq:moor}
\end{equation}
Note that this equation only differs from the Gaussian expression in \cref{eq:tostevin} by the additional factor 2 inside the square root.

The natural---but incorrect---expectation is that for large copy numbers both approximations converge to the true mutual information rate. However, the differences between \cref{eq:tostevin,eq:moor} already reveal that the two approximations do not converge. Indeed, 
 previous work has shown that even in the limit of infinite copy numbers, the Gaussian approximation only yields a lower bound which is not tight \cite{2023.Reinhardt,2023.Moor}. We further discuss this point below. 

\begin{figure}
    \centering
    \includegraphics[width=\linewidth]{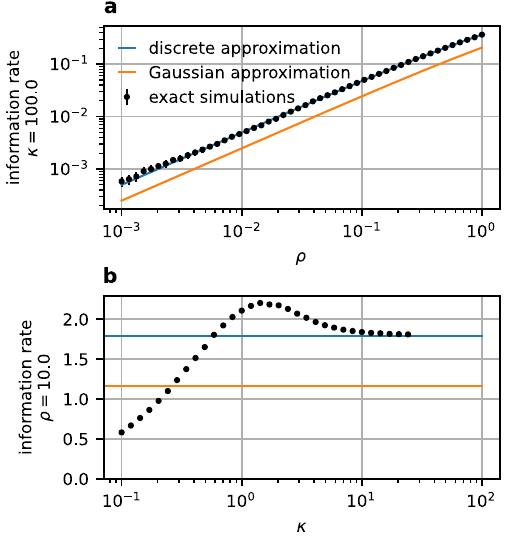}
    \caption{The mutual information rate of a simple linear reaction system defined by \cref{eq:reac1,eq:reac2,eq:reac3,eq:reac4}. The black dots show the exact information rate, computed with PWS. We compare both the Gaussian approximation of \citet{2009.Tostevin} and the discrete approximation of \citet{2023.Moor} against the exact result. In panel (a), we use parameters $\kappa=100$, $\lambda = 1$, $\mu = 1$ while varying $\rho$. The mean output copy number is directly proportional to $\rho$ with proportionality factor $\kappa\lambda^{-1}=100$. In panel (b), we fix $\rho=10$, $\lambda = 1$, $\mu = 1$, and systematically vary $\kappa$. As a consequence, we vary the mean input copy number $\bar{s}=\kappa\lambda^{-1}$, and simultaneously also the mean output copy number $\bar{x}=\bar{s}\rho\mu^{-1}=10\bar{s}$.
    }
    \label{fig:linear}
\end{figure}

We compared both approximations against exact PWS simulations for different parameter choices.
In \cref{fig:linear}a, we vary the mean copy number of the readout, $\bar{x}$, by varying its synthesis rate $\rho$ and compute the mutual information rate using both approximations as well as PWS, while keeping the input copy number constant at $\bar s = 100$.
We observe that the Gaussian approximation via the LNA [\cref{eq:tostevin}] consistently underestimates the mutual information rate. This confirms that even when $\bar{s}$ and $\bar{x}$ are large, the Gaussian approximation only yields a lower bound to the information rate of the discrete linear system. In contrast, the discrete approximation [\cref{eq:moor}] coincides with the true mutual information rate obtained from PWS simulations over all output copy numbers $\bar x$, even for $\bar x \ll 1$.

In \cref{fig:linear}b, instead of varying the copy number of the output, we vary the copy number of the input by varying the production rate $\kappa$.
Note that both the Gaussian approximation and the discrete approximation are independent of $\kappa$.
Yet, we observe that the true mutual information rate is not.
For sufficiently large input copy numbers the discrete approximation coincides with the true information rate while the Gaussian information rate remains only a lower bound. Thus, the discrete approximation is highly accurate for $\bar s \geq 10$. For small $\kappa$ where $\bar s < 10$, we find that the mutual information rate deviates from both the LNA as well as the discrete approximation. Surprisingly, we find an optimal value of $\kappa$ for which the mutual information rate is maximized and exceeds both approximations. This implies that at low input copy numbers, the system is able to extract additional information from the discrete input trajectories, which is not accounted for by either of the approximations. 

In all cases, we found that the Gaussian approximation deviates significantly from the true information rate for this discrete system.
Seemingly paradoxically, the Gaussian approximation based on the LNA does not converge to the true information rate at high copy numbers, even though the LNA approximates the stochastic dynamics extremely well in this regime.
In contrast, the discrete approximation from \citet{2023.Moor} does not suffer from this issue.
It has been shown that, generally, the Gaussian approximation is a lower bound on the discrete approximation, prompting the question of which features of the trajectories are not captured by the Gaussian approximation.

In recent work~\cite{2025.Moor}, we found that the root cause for the deviations of the Gaussian approximation lies in how the LNA approximates the reaction noise in the chemical master equation. 
While the dynamics of the chemical master equation give rise to discrete sample paths, i.e., piece-wise constant trajectories connected by instantaneous discontinuous jumps, the LNA approximation yields continuous stochastic trajectories.
Our results imply that a discrete sample path of $X$ carries more information about $S$ than the corresponding continuous sample path $x(t)$ would carry about $s(t)$ in the LNA.
In Ref.~\cite{2025.Moor} we show that this is ultimately due to the fact that in the discrete system, each reaction event is unambiguously recorded in the $X$ trajectories and thus different reactions modifying the same species can be distinguished. 
In contrast, in the continuous LNA description, all reactions that modify $X$ contribute to the noise term $\eta_x(t)$ in \cref{eq:output-dynamics} but their contributions are lumped together and 
therefore cannot be distinguished from an observed $x(t)$-trajectory. Specifically, note that for the motif studied here, only the production reaction $S\to S+X$ conveys information. The decay reaction of the output $X\to\emptyset$ does not carry information on the input fluctuations since its propensity is independent of the input. 
Yet, it contributes to the overall fluctuations in the output. 
The Gaussian approximation only considers the total fluctuations in the output, while the discrete approximation correctly distinguishes between the fluctuations induced from production events and decay events. 
Therefore, the Gaussian approximation consistently underestimates the true information transmission, whereas the discrete approximation does not incur this systematic error. This subtle point is reflected in the difference between \cref{eq:tostevin,eq:moor}.

\subsection{Nonlinear continuous system}
\label{sec:nonlinsys}

Next, we study a nonlinear variant of the reaction system above.
In contrast to the previous case study, we deliberately avoid using discrete dynamics, as we already observed that the Gaussian approximation is generally inaccurate in such systems. Instead, we focus solely on continuous Langevin dynamics to explore how an explicitly nonlinear input-output mapping affects the accuracy of the inherently linear Gaussian approximation. 
We hypothesize that the accuracy of the Gaussian approximation will deteriorate as the degree of nonlinearity increases. To test this hypothesis, we analyze a simple Langevin system with adjustable nonlinearity.

The system is defined by two coupled Langevin equations, one that describes the input, and one that describes the output.
The stochastic dynamics of the input $s(t)$ are given by \cref{eq:input-dynamics}. The output dynamics of $x(t)$ are given by
\begin{equation}
    \dot {x}(t) = \rho a(s) - \mu x(t) + \eta_x(t) \label{eq:nonlinear_x}
\end{equation}
with the Hill function
\begin{equation}
    a(s) =
    \begin{cases}
        \frac{s^n}{K^n + s^n} & \text{if } s \geq 0, \\
        0 & \text{if } s<0.
    \end{cases}
     \label{eq:hillfunc}
\end{equation}
This function serves as a tuneable non-linearity with the Hill coefficient $n$.
As $n\to0$, the Hill function approaches a shallow linear mapping, while for large $n$, it becomes sigmoidal and highly non-linear. As $n\to\infty$, $a(s)$ approaches the unit step function centered at $s=K$.
The so-called static input-output relation specifies the mean output $\bar{x}(s)$ for a static input signal $s$ and is given by $\bar{x}(s) = \rho a(s) / \mu$.
The gain of this system is then defined as the slope of this relation at $s=\bar{s}$, i.e., 
\begin{equation}
g= \frac{\partial \bar{x}(s)}{\partial s} \bigg|_{\bar{s}} = 
\frac{n a(\bar{s}) [1 - a(\bar{s})] \rho}{\mu\bar{s}} \,,
\label{eq:definition-gain}
\end{equation}
as derived in \cref{sec:gaussian-info-rate}. Importantly, for $\bar{s} = K$, the gain of the system is directly proportional to the Hill coefficient~$n$, i.e., the gain is directly coupled to the degree of nonlinearity.

\begin{figure}
    \centering
    \includegraphics{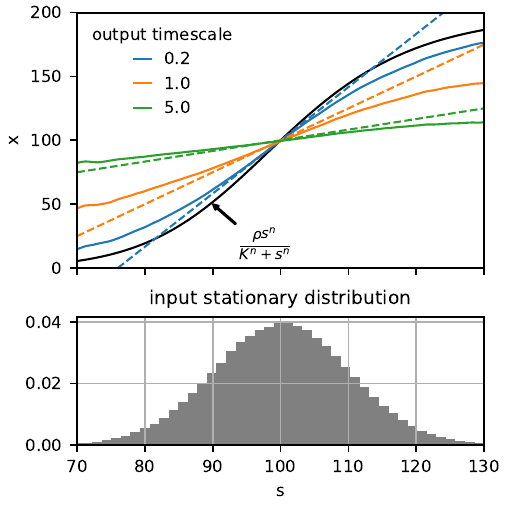}
    \caption{The dynamical input output relationship of the non-linear system with a fixed static gain of $g=5$ [\cref{eq:definition-gain}]. The upper panel shows the static input-output relationship (solid black line) as well as the dynamical input output relationship, i.e., the effective mapping from input to output $s(t)\mapsto x(t)$ (at the same time) for different output timescales $\tau_x = \mu^{-1}$. 
    The dynamical input-output mapping is defined as the conditional expectation $x_\text{dyn}(s)=\mathbb{E}[X(t) \mid S(t)=s]$ and was estimated non-parametrically from simulated trajectories of the system via Nadaraya–Watson kernel regression \cite{1964.Nadaraya,1964.Watson} with a Gaussian kernel (bandwidth~$h=\num{0.5}$).
   Additionally, using the linear noise approximation we obtain linear input output mappings with a dynamical gain $\tilde{g}=g/(1+\tau_x)$ (see \cref{sec:gaussian-info-rate}) which are displayed as dashed lines. 
    We observe that the linear mapping approximates the dynamical input output relation well for $s\approx\bar{s}=100$ but cannot capture the nonlinear saturation effect.
    The lower panel shows the stationary distribution of $s(t)$.}
    \label{fig:dynamical_input_output}
\end{figure}

The information rate quantifies the fidelity of the input–output mapping of the system and in dynamical systems this mapping is inherently limited by the response time scale.
\Cref{fig:dynamical_input_output} shows for various response time scales how the average output $\bar{x}(t)$ at a given time depends on the input $s(t)$ at that same time which is known as the dynamical input-output relation  \cite{malaguti_theory_2021}. The solid black curve represents the static input-output relation which is  determined by the instantaneous transfer function $a(s)$.
As the output responds more slowly to the input, the temporal input fluctuations are averaged out increasingly, leading to a shallower response for increasing $\tau_x$. Indeed, for very slow time scales, the system is well approximated by a linear responder.
Thus, we expect the Gaussian approximation of the information rate to be more accurate for slowly responding systems.

\begin{figure*}
    \centering
    \includegraphics{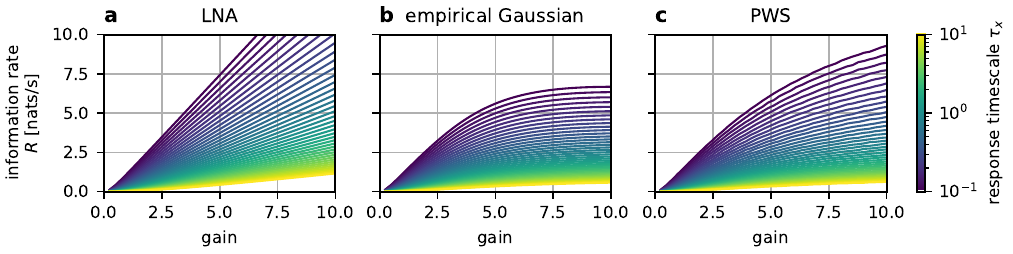}
    \caption{The information rate of a non-linear system as a function of its gain over a range of response timescales. 
    We vary the static gain $g$ by varying the Hill coefficient $n$, see \cref{eq:definition-gain}.
    A short response timescale corresponds to a fast system (purple) while a long response timescale corresponds to a slow system (yellow). The information rate was computed in three different ways: (\textbf{a}) via the Gaussian approximation using the LNA to estimate the required power spectra; (\textbf{b}) via the Gaussian approximation using simulations to numerically estimate the required power spectra; (\textbf{c}) exactly via PWS.}
    \label{fig:nonlinear}
\end{figure*}

With the Langevin extension of PWS we can directly compute the mutual information rate of this nonlinear model whereas the Gaussian approximation can only be applied to linear systems. 
Therefore, to obtain the mutual information rate in the Gaussian approximation, we have to linearize the system.
There are two approaches for linearizing the stochastic dynamics of this nonlinear system which result in different information estimates.

The first approach is to linearize \cref{eq:nonlinear_x} analytically via the LNA as shown in \cref{app:gauss:lna}. Within this approach, we can obtain an analytical expression for the information rate (see \cref{sec:gaussian-info-rate}),
\begin{equation}
    R_\text{LNA} = \frac{\lambda}{2}\left(\sqrt{1+g^2 \frac{\bar s \mu}{\bar x \lambda}}-1\right).
\end{equation}
This LNA-based approach also yields a linearized dynamic input-output relation, shown as dashed lines in \cref{fig:dynamical_input_output}.

We observe that the linearized input-output relation closely matches the slope of the true nonlinear dynamical input-output relation at $s=\bar s = 100$, but overall it does not correspond to a (least-squares) linear fit of the nonlinear dynamical input-output relation.
For all values of $s$, the linearized input-output relation has a slope greater than or equal to the slope of the dynamical input-output relation.
Empirically, the LNA thus seems to over-estimate the dynamic gain of the system.
The reason may be that the LNA approximates the static input-relation (\cref{fig:dynamical_input_output} black curve), and estimates the linearized dynamical input-output relation based on this static approximation only.

The second ``empirical Gaussian'' approach to linearize the nonlinear system avoids these issues.
In this approach, we first numerically generate trajectories from the stochastic \cref{eq:input-dynamics,eq:nonlinear_x} and use digital signal processing techniques to estimate the mutual information rate from the trajectories.
We numerically estimate the (cross) power spectra of input and response using Welch's method \cite[see Ch.~11]{1975.Oppenheim}. 
From the estimated spectral densities $\hat{S}_{\alpha\beta}(\omega)$ we compute the coherence
\begin{equation}
    \hat\phi_{sx}(\omega) = \frac{|\hat S_{sx}(\omega)|^2}{\hat S_{ss}(\omega) \hat S_{xx}(\omega)}
\end{equation}
which we use to obtain the Gaussian approximation of the mutual information rate directly using \cref{eq:gaussdef}.

The empirical power spectra characterize the linear response of a system, but not in the same way as the LNA.
While for linear systems the power spectra obtained via the LNA match the empirical power spectra \cite{2006.Warren}, for a nonlinear system, the empirical power spectra and the coherence can differ from the corresponding LNA calculations. 
The two linearization approaches are thus not equivalent.
We tested the accuracy of the Gaussian mutual information rate estimates using both linearization approaches to elucidate the differences in these approaches.

\begin{figure}
    \centering
    \includegraphics{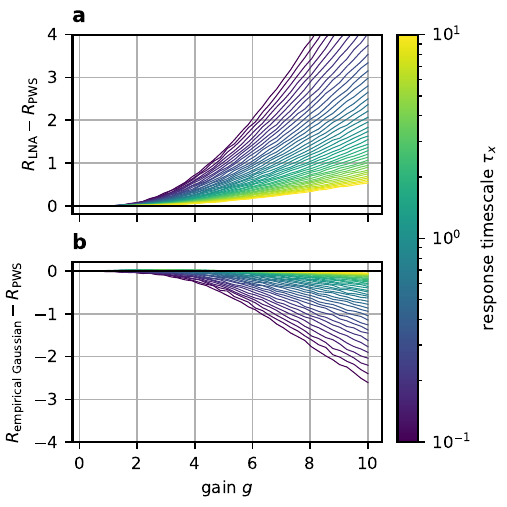}
    \caption{Deviation from the exact information rate for the approximate Gaussian information rate computed via LNA (top), and the approximate Gaussian information rate computed via empirical estimates of the power spectra (bottom). 
    A deviation of 0 implies perfect accuracy. 
    While the absolute deviation of both approximations increases with increasing gain and decreasing response timescale, the LNA based approach consistently overestimates the information rate whereas the empirical approach constitutes a lower bound. Moreover, in terms of absolute deviation, the empirical approach is more accurate across all parameter values.}
    \label{fig:lna_pws_absolute}
\end{figure}

\Cref{fig:nonlinear} displays the mutual information rate obtained via two linearized approximations as well as the exact PWS result.
We vary the gain $g$ and the response time-scale $\tau_x$, both of which significantly affect the shape of the dynamical input-output relationship.
As expected, a larger gain or a faster response time lead to an increase in the mutual information rate.
At large gain, the information rate naturally saturates as $a(s)$ approaches a step function.
The saturation effect is clearly seen in the PWS results, and is found to be even more pronounced in the empirical Gaussian approximation.
The LNA-based Gaussian approximation, however, shows no saturation.
This highlights that the LNA directly linearizes the system at the level of the input-output mapping $a(s)$ which results in an approximation that is unaffected by the sigmoidal shape of $a(s)$.
In contrast, the empirical approximation is affected by nonlinear saturation effects because it is computed from simulated trajectories.
We thus see that both approximations yield substantially different results at large gain.

In \cref{fig:lna_pws_absolute} we compare the absolute deviation between the approximations and the PWS result.
For small gain we see that both approximations are accurate which is not surprising since the nonlinearity is very weak in this regime.
For large gain, the LNA-based approximation always overestimates the mutual information rate while the empirical Gaussian approximation always underestimates the rate.
In both cases the systematic error decreases as the response timescale becomes slower. This reflects the fact that for slow responders, the dynamic input-output relation is more linear (\cref{fig:dynamical_input_output}) than for fast responders.

Additionally, we computed the relative deviation, see \cref{fig:lna_pws_ratio} in \cref{sec:relative-deviation}. We find that in terms of relative error the curves for different response timescales largely overlap. In terms of relative approximation error, the gain, rather than response timescale, is the primary factor affecting the accuracy of the Gaussian approximation.

\section{Discussion}

We investigated the accuracy of the Gaussian approximation for the mutual information rate in two case studies, each highlighting a scenario where the approximation may be inaccurate.
We were able to reliably quantify the inaccuracy in each case by comparing against the ``ground truth'' mutual information rate for these scenarios which was computed using a recently developed exact Monte Carlo technique called PWS \cite{2023.Reinhardt}.

We first considered linear discrete systems, which are relevant in biology due to the discrete nature of biochemical signaling networks. In our example, the Gaussian approximation cannot capture the full information rate, but only yields a lower bound. We show that a discrete approximation, developed by \citet{2023.Moor}, is able to correctly estimate the mutual information rate of the network over a wide range of parameters. 
Since the Gaussian approximation captures the second moments of the discrete system,
this finding demonstrates that a discrete system can transmit significantly more information than what would be inferred from its second moments alone.
This perhaps surprising fact has been highlighted before~\cite{2019.Cepeda-Humerez,2023.Moor,2023.Reinhardt}, and it hinges on the use of a discrete reaction-based readout.
As demonstrated in Ref.~\cite{2025.Moor}, 
the increased mutual information rate found for a discrete readout stems from the ability of unambiguously distinguishing between individual reaction events in the readout's trajectory.
However, it remains an open question whether biological (or other) signaling systems can effectively harness this additional information encoded in the discrete reaction events.
For systems that cannot distinguish individual reaction events in downstream processing, the Gaussian framework might still accurately quantify the ``accessible information''.

A notable new observation in our first case study is the deviation between the discrete approximation of the mutual information rate derived by \citet{2023.Moor} and the exact result obtained using PWS \cite{2023.Reinhardt} for inputs with low copy number $\bar s$.
In the discrete approximation, the mutual information rate is independent of the input copy number $\bar s$, but the PWS simulations show that at low copy numbers there is an optimal $s^\star \approx 1$ which maximizes the mutual information rate, see \cref{fig:linear}b.
This surprising finding suggests that the information rate in discrete systems can be increased by reducing the copy number of the input sufficiently, such that it only switches between a few discrete input levels.
Notably, in the reverse case---low output copy number $\bar x$ but large $\bar s$---the discrete approximation always remains accurate, see \cref{fig:linear}a. We leave a precise characterization of this finding for future work.

The second case study focused on a continuous but nonlinear system, where we demonstrated that the accuracy of the Gaussian approximation depends on the linearization method. Linearizing the underlying system dynamics directly via the LNA leads to an overestimation of the information rate, while estimating the system's correlation functions empirically from data underestimates it. Regardless of the method, the Gaussian approximation is more accurate in terms of absolute deviation of the true information rate when the gain of the system is small and its response is slow compared to the timescale of the input fluctuations.

The result of our second case study---that the empirical Gaussian mutual information rate underestimates the true rate---is consistent with theoretical expectations.
As shown by \citet{mitra2001nonlinear}, and highlighted in \cite{2010.Tostevin}, an empirical Gaussian estimate of the mutual information between a Gaussian input signal $S_G$ and a non-Gaussian output $X$ provides a lower bound on the channel capacity $C(S,X)=\max_{\prob(s)} I(S, X)$ (subject to a power constraint on $S$). 
Specifically, they show that $C(S, X) \geq I(S_G; X) \geq I(S_G; X_G)$, where $(S_G, X_G)$ is a jointly Gaussian pair with the same covariance matrix as $(S_G, X)$.
For purely Gaussian systems like $(S_G, X_G)$, the mutual information calculated using \cref{eq:gaussdef} is exact and equal to the channel capacity. 
However, for systems that have a Gaussian input but are otherwise non-Gaussian, the mutual information is larger or equal than the corresponding Gaussian model with matching second moments, as evidenced in \cref{fig:lna_pws_absolute}.
In general, the empirical Gaussian approximation yields a lower bound on the mutual information of the nonlinear system with a Gaussian input signal,
as well as a lower bound on the channel capacity of the nonlinear system
\footnote{Note that this argument does not apply to the Linear Noise Approximation (LNA). The bound specifically requires the Gaussian model to use the covariance of the full, original system. When the system is first linearized using the LNA, the resulting linear model does not retain the same covariance as the original nonlinear system. As a result, the mutual information rate calculated with the LNA is not a lower (nor an upper) bound on the true mutual information rate.}.

We can distill several concrete recommendations for the computation of the information rate from our analysis. For linear discrete systems, the Gaussian approximation yields a lower bound on the true information rate which may accurately quantify the information available to systems that cannot distinguish individual discrete events. Alternatively, the reaction-based discrete approximation by \citet{2023.Moor} is highly accurate, even when the copy number of the output is extremely small. However, when the copy number of the input becomes small ($\lesssim 10$), both approximations break down and one must use an exact method. 
Exact methods for obtaining the information rate of any stochastic reaction-based systems are PWS \cite{2023.Reinhardt} or brute-force numerical integration of the stochastic filtering equation, as shown in \cite{2023.Moor}.
For nonlinear continuous systems with small gain one can safely use the Gaussian approximation, either based on a linearization of the underlying dynamics or on empirically estimated correlation functions.  
Moreover, when the slowest input time-scale is more than a magnitude faster than the response time-scale, the non-linear response of the system is averaged out by the quick input fluctuations, and the Gaussian approximation yields accurate results. 
In this case, using a Gaussian approximation based on empirical correlation functions yields the most accurate result, and provides a lower bound for the mutual information. 
Finally, if the system is both highly nonlinear and has a fast response with respect to the input, one must resort to an exact method like PWS. 
We hope that our results will guide future research in determining the appropriate method for computing the mutual information rate.

Overall, our results greatly increase the usefulness of the Gaussian approximation for the information rate of non-Gaussian systems. 
The Gaussian approximation remains a useful method that can be applied directly and straightforwardly to experimental data. 
Here, we have quantified the prerequisites to safely use this approach. 
Moreover, we elucidate how an empirical Gaussian approximation constitutes a lower bound on the true information rate for systems with a sufficiently large input copy number.

\begin{acknowledgments}
This work is part of the Dutch Research Council (NWO) and was performed at the research institute AMOLF.
This project has received funding from the European Research Council (ERC) under the European Union’s Horizon 2020 research and innovation program (grant agreement No.~885065), 
and was financially supported by NWO through the “Building a Synthetic Cell (BaSyC)” Gravitation grant (024.003.019).
\end{acknowledgments}

\appendix
\setcounter{figure}{0}
\renewcommand{\thefigure}{S\arabic{figure}}

\section{Gaussian approximation}
\label{app:gauss}
Here we derive the analytical expressions for the Gaussian information rate of the networks considered in the main text. To this end we first discuss the dynamics of the input signal $S$ and its power spectrum. Then, we perform a linear approximation of the dynamics of the readout species $X$ and derive the approximate Gaussian information rate between $S$ and $X$ for the nonlinear network. Finally, we derive the Gaussian information rate of the linear network from our expression of the Gaussian information rate of the nonlinear network.

\subsection{Signal}
The input signal is generated by a birth-death process,
\begin{equation}
\label{eq:SBD}
\begin{aligned}
    \ce{\emptyset &<=>[\kappa][\lambda] S},\\
\end{aligned}
\end{equation}
Its dynamics in Langevin form are,
\begin{equation}
    \dot{s} = \kappa -  \lambda s(t)+ \eta_s(t), \label{eq:sigdyn}
\end{equation}
yielding the steady state signal concentration $\bar s=\kappa/\lambda$. The independent Gaussian white noise process $\eta_s(t)$ summarizes all reactions that contribute to fluctuations in $S$. The strength of the noise term in steady state is 
\begin{equation}
    \avg{\eta_s^2} = \kappa+ \lambda\bar s=2\lambda \bar s. \label{eq:etas}
\end{equation}

The power spectral density, or power spectrum, of a stationary process $\mathcal{X}$ is defined as $\spec{xx} = \lim_{T \to \infty}\frac{1}{T} |\f{x_T}|^2$, where $\f{x}$ denotes the Fourier transform of $x(t)$. The power spectrum of a signal obeying \cref{eq:sigdyn} is thus given by
\begin{equation}
\label{eq:specss}
\begin{aligned}
    \spec{ss} 
    &= \frac{\avg{\eta_s^2}}{\omega^2+\lambda^2}=\frac{2 \lambda \bar s}{\omega^2+\lambda^2}. 
\end{aligned}
\end{equation}

\subsection{Linear approximation}
\label{app:gauss:lna}
We now consider the readout $X$, which is produced via a nonlinear activation function $a(s)$:
\begin{equation}
\label{eq:StoV}
\begin{aligned}
    \ce{S &->[\rho \,a(s)] S + X},\\
    \ce{X &->[\mu] \emptyset}.
\end{aligned}
\end{equation}
We define the activation level $a(s)$ to be a Hill function,
\begin{equation}
\label{eq:as}
	a(s) = \frac{s(t)^n}{K^n + s(t)^n}.
\end{equation}
Such a dependency, in which $K$ sets the concentration of $S$ at which the activation is half-maximal and $n$ sets the steepness, can for example arise from cooperativity between the signal molecules in activating the synthesis of $X$. 

We have for the dynamics of $X$ in Langevin form
\begin{equation}
	\dot{x} =  \rho \, a(s) - \mu \, x(t) + \eta_x(t),
\end{equation}
with $a(s)$ given by \cref{eq:as}. The steady state concentration of $X$ is given by $\bar x = \bar a \rho/\mu$, where we have defined the steady state activation level $\bar a = a(\bar s)$. It is useful to determine the static gain of the network, which is defined as the change in the steady state of the output upon a change in the steady state of the signal:
\begin{equation}
\label{eq:g}
\begin{aligned}
	g &= \partial_{\bar s} \bar x = r/\mu, \\ 
	& = n(1-\bar a) \bar x/\bar s,
\end{aligned}
\end{equation}
where we have defined the approximate linear activation rate
\begin{equation}
\label{eq:r}
	r =  n \bar a(1-\bar a)\rho/\bar s,
\end{equation}
and the steady state of the activation level is given by
\begin{equation}
	\bar a =\frac{\bar s^n}{K^n + \bar s^n}.
\end{equation}
Generally, we assume that $K=\bar s$, which entails that in steady state the network is tuned to $\bar a = 1/2$. 

To compute the Gaussian information rate we approximate the dynamics of $X$ to first order around $\bar x$ via the classical linear noise approximation \cite{vanKampen1992}. Within this approximation the dynamics of the deviation $\delta x(t) = x(t) -\bar x$ are,
\begin{equation}
\label{eq:xlindyn}
	\dot{\delta x} = r\,  \delta s(t)- \mu \, \delta x(t) + \eta_x(t),
\end{equation}
with the synthesis rate $r$ given by \cref{eq:r}. 

In the linear noise approximation the noise strength is a constant given by the noise strength at steady state,
\begin{equation}
\label{eq:etax}
	\avg{\eta_x^2} = \rho \bar a + \mu \bar x = 2 \mu \bar x. 
\end{equation}

\subsection{\label{sec:gaussian-info-rate}Information rate}
Following Tostevin \& Ten Wolde \cite{2009.Tostevin,2010.Tostevin}, we can express the Gaussian information rate as follows,
\begin{equation}
\label{eq:infratedef}
	R(\mathcal S; \mathcal X) =  \frac{1}{4\pi}\int_{-\infty}^{\infty} d\omega \log \left(1+ \frac{\freqgain}{\freqnoise}\spec{ss}\right),
\end{equation}
where $\freqgain$ is the frequency dependent gain and $\freqnoise$ is the frequency dependent noise of the output process $\mathcal X$. If the intrinsic noise of the network is not correlated to the process that drives the signal, the power spectrum of the network output obeys the spectral addition rule \cite{tanase2006signal}. In this case the frequency dependent gain and noise can be identified directly from the power spectrum of the output, because it takes the following form:
\begin{equation}
\label{eq:specxx}
    \spec{xx} = \freqgain \spec{ss} + \freqnoise.
\end{equation}
For a species $X$ obeying \cref{eq:xlindyn}, we have
\begin{equation}
\label{eq:fgfn}
    \begin{aligned}
    \freqgain &= \frac{r^2}{\mu^2 + \omega^2}, \\
 	\freqnoise &= \frac{\avg{\eta_x^2}}{\mu^2 + \omega^2}= \frac{2 \mu \bar x}{\mu^2 + \omega^2}.
    \end{aligned}
\end{equation}
The Wiener Khinchin theorem states that the power spectrum of a stochastic process and its auto-correlation function are a Fourier transform pair. We thus obtain for the variance in the readout, substituting the frequency dependent gain and noise [\cref{eq:fgfn}] and the power spectrum of the signal [\cref{eq:specss}] in \cref{eq:specxx} and taking the inverse Fourier transform at $t=0$,
\begin{equation}
\label{eq:varx}
    \sigma_x^2 = g\dg\sigma_s^2 + \sigma^2_{x|s}=\frac{g^2 \bar s}{1+\lambda/\mu} + \bar x,
\end{equation}
where the signal variance equals its mean $\sigma_s^2=\bar s$, and the mean readout concentration sets the intrinsic noise $\sigma^2_{x|s}=\bar x$. We further have the static gain $g$ given by \cref{eq:g}, and have defined the dynamic gain 
\begin{equation}
\label{eq:dg}
    \dg \equiv \frac{\avg{\delta x(t) \delta s(t)}}{\sigma_s^2}= \frac{g}{1+\lambda/\mu},
\end{equation}
which is the slope of the mapping from the time-varying signal value $s(t)$ to the time-varying readout $x(t)$; for Gaussian systems $\avg{x(t)|s(t)} = \dg s(t)$ \cite{2010.Tostevin,malaguti_theory_2021,tjalma2024predicting}. 
 
To solve the integral in \cref{eq:infratedef} we exploit that
\begin{equation}
\label{eq:infrate1root}
	\int^\infty_{-\infty} d\omega \log\left( \frac{\omega^2 + a^2}{\omega^2 + b^2}\right) = 2\pi(a-b). 
\end{equation}
Substituting the frequency dependent gain and noise given in \cref{eq:fgfn} and the signal power spectrum of \cref{eq:specss} in \cref{eq:infratedef} and using \cref{eq:infrate1root} we obtain the information rate,
\begin{equation}
\label{eq:Rnl}
    \begin{aligned}
        R(\mathcal S; \mathcal X) &= \frac{\lambda}{2}\left(\sqrt{1+\frac{r^2\avg{\eta_s^2}}{\lambda^2\avg{\eta_x^2}}}-1\right), \\
        &= \frac{\lambda}{2}\left(\sqrt{1+g^2\frac{\bar s \mu}{\bar x \lambda}}-1\right),
    \end{aligned}
\end{equation}
where we used the noise strengths given in \cref{eq:etas} and \cref{eq:etax}, we have the static gain $g$ of \cref{eq:g}, and the synthesis rate $r$ of \cref{eq:r}. 


\subsection{Linear network}
To disambiguate differences in the information rate caused by the linear approximation of our nonlinear reaction network on the one hand and the Gaussian approximation of the underlying jump process on the other, we consider the information rate of a linear network. Any difference between the exact information rate and the Gaussian information rate must then be a result of the Gaussian approximation. To this end we use the same input signal [\cref{eq:sigdyn}], but we consider a linear activation of the readout, i.e.
\begin{equation}
\begin{aligned}
    \ce{S &->[\rho] S + X},\\
    \ce{X &->[\mu] \emptyset}
\end{aligned}
\end{equation}
such that the Langevin dynamics of $X$ are
\begin{equation}
    \dot x = \rho s(t) - \mu x(t) + \eta_x(t),
\end{equation}
which yields the steady state concentration $\bar x= \bar s \rho/\mu$. For this linear readout, the static gain is simply set by the ratio of steady states of the input and the output, $g=\rho/\mu=\bar x/\bar s$. We can then obtain the information rate of this linear system by substitution of its static gain in \cref{eq:Rnl}, which yields
\begin{equation}
\begin{aligned}
\label{eq:Rlin}
        R(\mathcal S; \mathcal X) 
        & = \frac{\lambda}{2}\left(\sqrt{1+\frac{\rho}{\lambda}}-1\right).
        \end{aligned}
\end{equation}
This result is identical to that of \citet{2009.Tostevin, 2010.Tostevin} (motif III). 

\section{Relative deviation of the Gaussian approximation for a nonlinear system}
\label{sec:relative-deviation}

\begin{figure}
    \centering
    \includegraphics{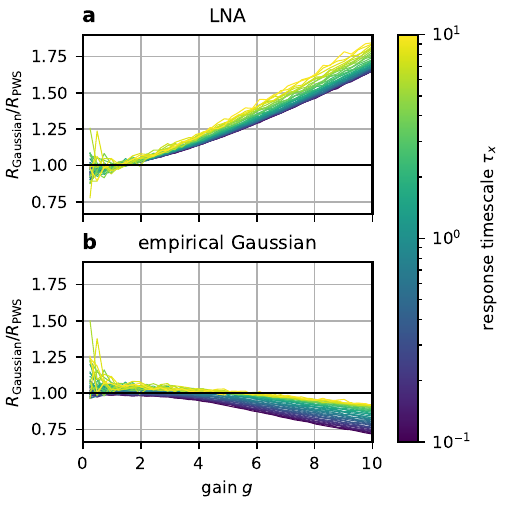}
    \caption{Relative deviation from the exact information rate for the approximate Gaussian information rate. The relative deviation was computed for both the LNA-based approximation and the empirical Gaussian approximation using numerically estimated power spectra, across varying system gains and response timescales (see \cref{sec:nonlinsys} for details). The curves for different response timescales largely overlap, indicating that the relative approximation error is primarily influenced by system gain rather than response timescale. This highlights that system gain is the dominant factor in determining the accuracy of the Gaussian approximation.}
    \label{fig:lna_pws_ratio}
\end{figure}

Due to the definition of the mutual information, an absolute difference in information maps to a relative difference in the reduction of uncertainty.
Therefore, \cref{fig:lna_pws_absolute} in \cref{sec:nonlinsys} focuses on the absolute deviation between the Gaussian approximation and the true mutual information.
Nevertheless, the relative deviation of the Gaussian information rate from the true rate can still provide valuable insights, and we discuss it here.

In \cref{fig:lna_pws_ratio} we compare the relative deviation $R_\text{Gaussian} / R_\text{PWS}$ between the Gaussian approximation and the exact mutual information computed using PWS.
We find that the relative deviation increases as the system gain increases, indicating that the Gaussian approximation also becomes relatively less accurate for larger gains.
As already discussed above, the empirical Gaussian method consistently underestimates the true information rate, while the LNA-based approximation overestimates it.

Interestingly, we also observe that for the LNA approximation, at fast timescales the result is slightly more accurate, whereas the empirical Gaussian estimate is more accurate at slow timescales. We initially expected that in both cases slow timescales would yield better agreement with PWS, as the input-output dynamics are more linear for slow timescales, and thus better approximated by the Gaussian model. The fact that this is not the case for the LNA approximation is intriguing, indicating the need for further investigation into the interplay between timescales, system nonlinearity, and the LNA.

\bibliography{library}

\end{document}